# Impact of Extended Reality on Robot-Assisted Surgery Training

**Michael Bickford[1], Fayez Alruwaili[2], Sara Ragab[1], Hanna Rothenberg[1], and Mohammad Abedin-Nasab[2,*]**

[1]Rowan-Virtua School of Osteopathic Medicine, Rowan University, Stratford, NJ, 08084, United States
[2]Department of Biomedical Engineering, Rowan University, Glassboro, NJ, 08028, United States

[*]E-mail: abedin@rowan.edu

**Abstract**

Robot Assisted Surgeries (RAS) have one of the steepest learning curves of any type of surgery. Because of this, methods to practice RAS outside the operating room have been developed to improve the surgeons' skills. These strategies include the incorporation of extended reality simulators into surgical training programs. In this Systematic review, we seek to determine if extended reality simulators can improve the performance of novice surgeons and how their performance compares to the conventional training of surgeons on Surgical robots. Using the PRISMA 2020 guidelines, a systematic review and meta-analysis was performed searching PubMed, Embase, Web of Science, and Cochrane library for studies that compared the performance of novice surgeons that received no additional training, trained with extended reality, or trained with inanimate physical simulators (conventional additional training). We included articles that gauged performance using either GEARS or Time to complete measurements and used SPSS to perform a meta-analysis to compare the performance outcomes of the surgeons after training. Surgeons trained using extended reality completed their surgical tasks statistically significantly faster than those who did not receive training (Cohen's d=-0.95, p=0.02), and moderately slower than those conventionally trained (Cohen's d=0.65, p=0.14). However, this difference was not statistically significant. Surgeons trained on extended reality demonstrated a statistically significant improvement in GEARS scores over those who did not train (Cohen's d=0.964, p<0.001). While surgeons trained in extended reality had comparable GEARS scores to surgeons trained conventionally (Cohen's d=0.65, p=0.14). This meta-analysis demonstrates that extended reality simulators translated complex skills to surgeons in a low cost and low risk environment. Highlighting the value of incorporating innovative training regimens into surgical practice. This study is the first to highlight the positive impact extended reality can have on preparing surgeons for robotic assisted surgery compared to traditional training.

# 1. Introduction

Robot-assisted surgery (RAS) has become the standard practice for many surgical procedures, revolutionizing the field with its enhanced precision, control, and minimally invasive techniques [1,2]. As robotic surgery develops, more procedures originally done manually are now being performed using surgical robots [3]. However, mastering the skills of robotic systems presents significant challenges due to the steep learning curve and extensive training required before being able to perform the surgeries effectively [4], [5].

RAS demands not only traditional surgical skills but also proficiency in operating complex robotic systems. Acquiring these skills involves extensive training to maneuver instruments safely and effectively [4], [5]. The unique challenges of manipulating robotic instruments through small incisions, often with limited tactile feedback, and the need to coordinate hand-eye movements while operating through a remote interface adds many layers of complexity on top of performing the surgery itself. Learning these skills solely in the operating room can be inefficient, time-consuming, and pose safety concerns for patients [4], [5]

Acquiring both technical and non-technical skills outside the operating room is now a crucial component of RAS training. Beyond traditional methods, including dry, wet, and cadaveric labs, extended reality (XR) simulators have gained significant importance in the early stages of technical skill development for RAS [6], [7]. XR simulators are safe, ethical, and repeatable alternatives that produce objective measures of performance and allow real-time feedback to trainees, often without the need for constant supervision of an experienced surgeon in a safe environment.

The emergence of XR, which encompasses virtual reality (VR), augmented reality (AR), and mixed reality (MR), offers further advancements in surgical training [8], [9]. These technologies enable trainees to practice surgical skills on a limitless number of cases with real-time feedback, thus enhancing learning outcomes and skill acquisition. Many simulators have been produced and are available in the market, giving greater access to novice surgeons (Recent literature underscores the growing importance of XR in surgical training. For instance, a 2023 study by Co et al. evaluate the XR-based systems for surgical education [9]. This study emphasizes the potential of XR to revolutionize surgical education by offering unlimited practice opportunities and real-time feedback, thereby bridging the gap between theoretical knowledge and practical skills.

Despite the growing adoption of XR-based training, no systematic review has evaluated XR training programs to other traditional methods specifically for RAS. In this systematic review, we evaluate the XR training program compared with traditional training methods, aiming to provide insights into its effectiveness in preparing surgical trainees for the complexities of RAS. This comparison will provide valuable information on how best to integrate these emerging technologies into RAS training and improve the proficiency and safety of future surgeons.

# 2. Method

A systematic review and meta-analysis of randomized control trials and prospective studies was conducted using the PRISMA 2020 guidelines.

## 2.1 Search strategy

4 databases (PubMed, Embase, Web of Science, and Cochrane library) were searched on April 14$^{th}$, 2024 using the search string ("Robot assisted Surgery" OR "Robotic Surgery" OR "Robotic Surgical") AND ("Training" OR "Program" OR "Instruction" OR "Teach") AND ("Virtual reality" OR "Mixed Reality" OR "Augmented Reality" OR "VR").

## 2.2 Study selection

The search resulted in 1,267 articles that were exported to Rayyan.ai for duplication detection (Figure 1). After screening for duplicates, 2 authors independently screened 700 articles by title and abstract based on the studies inclusion and exclusion criteria. This left 36 articles for full text review. Each article was screened independently by 2 authors and then finalized into 14 articles.

*2.3 Inclusion and exclusion criteria*

Articles included in this paper were randomized controlled trials and prospective studies that compared inexperienced surgeons trained with an extended reality training program for robot assisted surgery to inexperienced surgeons trained with on a robotic surgical device with an inanimate model referred to as a dry lab, or inexperienced surgeons with no additional training. In particular, the authors looked for papers that measured performance of the surgeons before and after the training with quantitative results like GEARS and Time to complete. Excluded articles included virtual reality validation studies where all participants used the simulator, abstracts, review articles, and articles that did not have GEARS or time to complete.

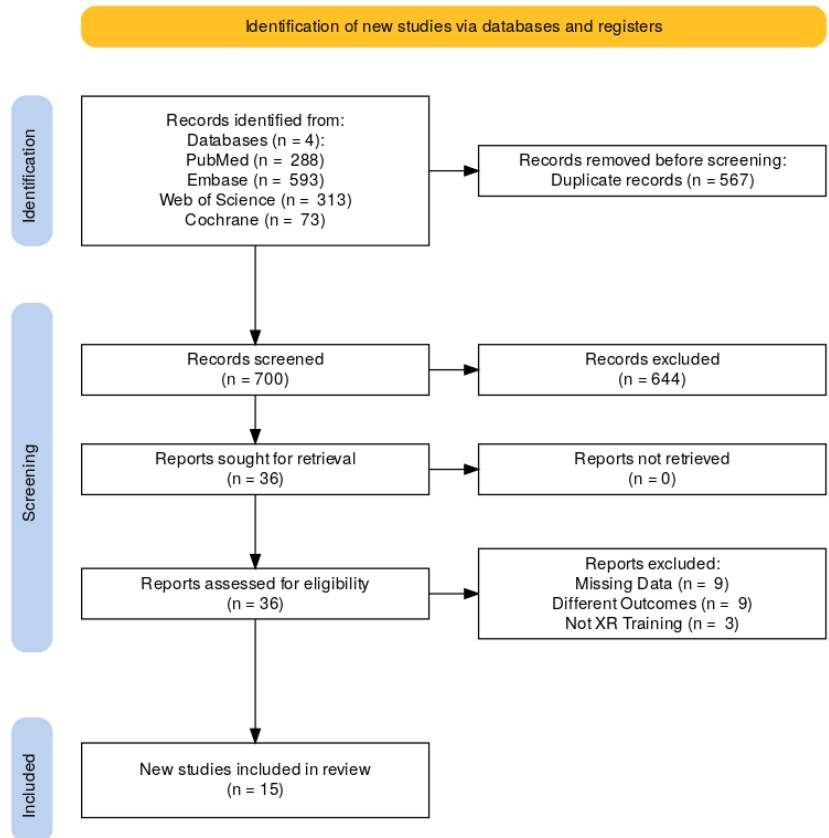

**Figure 1.** PRISMA flow diagram of the systematic review

*2.4 Data extraction*

Qualitative data was extracted from the articles after the study selection was complete. The Global Evaluative Assessment of Robotic Skills (GEARS) and time to complete were the primary outcomes observed in this paper. GEARS is a tool used to evaluate a participant's ability to perform robotic surgical procedures [10]. Surgical performance is usually assessed by an expert surgeon in 6 domains related to the fundamental robotic surgical skills such as depth perception. Each domain is scored on a 5 point anchored Likert-scale with a final score given as a summation of all 6 domains. Time to complete is a measure of how long each participant took to complete a predefined task, in this paper, we looked specifically into suturing and knot tying with time measured or converted to seconds.

*2.5 Statistical analysis*

To analyze time to complete; the mean, Standard deviation, and number of participants were collected before and after the training program was completed. A meta-analysis was performed using a random effects model on IBM SPSS Statistics for Windows, version 29 (IBM Corp., Armonk, N.Y., USA). The meta-analysis pooled the effect sizes of each article to evaluate the change in means in time to complete before and after the training program. The extent of improvement in scores was determined by the effect size of the analysis (Cohen's d) with 95% confidence intervals. The variance between each study was calculated using a heterogeneity test using Q statistics and the $I^2$ ratio ($I^2 = \tau^2/H^2$). A larger $I^2$ value suggests a higher level of variance between studies. The authors compared the gear scores of participants post training by using an independent sample t-test with a p value of <0.05 determining statistical significance.

*2.6 Risk of Bias*

Each article was independently reviewed for risk of bias and quality of evidence by 2 authors. Risk of bias was determined using the Cochrane Risk of bias template. The quality of evidence was determined using the Grading of Recommendations Assessment, Development and Evaluation (GRADE). Because this article included both Randomized control trials and prospective studies, Robbins-I was used to evaluate each article.

## 3. Results
*3.1 Summary of Findings*

A total of 1,267 articles were generated through electronic search. After deletion of duplicates and inclusion and exclusion criteria were applied, 14 randomized control trials and prospective studies were included in this study with a total of 587 participants. Each article included in the analysis is described in the summary of articles table (Table 1). 13 articles trained participants in virtual reality training while Chowriappa et al, 2014 looked at augmented reality training. 8 of the included articles compared virtual reality training to no additional training. 8 articles compared extended reality training to Dry lab training. There is an overlap with the articles Valdis et al. and Kortes et al. comparing all 3 categories of participants and were therefore included in both analyses. The participants in these studies ranged from medical students to attending surgeons, however, all had limited experience when it came to robot assisted surgeries. The two simulators used by most studies were the da Vinci Surgical Simulator (dVSS) and the Mimic dV-Trainer.

*3.2 Summary of Surgical simulators*

Many surgical simulators have been developed and are available to be used to train surgeons. The most popular of these simulators is the dVSS, which is useful for training surgeons in any robotic assisted surgery and prepares surgeons for using the da Vinci surgical system [11]. Many of the other simulators fill similar roles, or for specific styles of surgery not covered by the dVSS and are either virtual reality, augmented reality, or a hybrid of both (Table 2).

*3.3 XR vs. No Additional Training*

To determine if virtual reality is an effective form of training surgeons. We compared participants that were trained in virtual reality to participants that did not receive any training outside of being familiar with the surgical robot. The time to complete and GEARS score of a final trial after the training program were used to compare the two groups.

A random effects analysis was used for this meta-analysis and showed that participants with virtual reality training took statistically significant less time to complete tasks than participants with no additional training (Cohen's d=-0.95, 95% CI [-1.73, -0.16], p=0.02) (Figure 2).

**Table 1. Summary Table of Included Studies**

| STUDY ID (FIRST AUTHOR, YEAR) | STUDY DESIGN | TYPE OF PARTICIPANTS | # OF PARTICIPANTS | VR SIMULATOR | COMPARATOR | ASSESSMENT METHOD | INCLUSION CRITERIA | OUTCOMES COLLECTED | OUTCOMES MEASURED |
|---|---|---|---|---|---|---|---|---|---|
| Kortes, 2011 | Randomized Controlled Trial | Residents | 16 | MdVT | da Vinci Surgical System dry lab and no additional | Copper wire transfer and Suturing exercises using the da Vinci Surgical System | Urology residents with novice to intermediate levels of training | Baseline evaluation before training and secondary evaluation | Time to complete, OSATS |
| Raison, 2020 | Prospective Study | Medical students | 43 | dVSS | da Vinci Surgical System dry lab | Assessed during last 10 minutes of every training session and final assessment | Novice surgeons with limited robot or simulation experinence | Collected after each of the 3 training sessions | GEARS |
| Kiely, 2015 | Randomized Controlled Trial | Residents and Attendings | 23 | dVSS | No additional Training | Before and After assessment on an inanimate vaginal cuff model | Residents and attendings from the department of OBGYN, Urology, General Surgery, or Cardiac | Videos of assessments recorded and sent to 3 blind evaluators | GOALS+, GEARS, Total knots, Satisfactory knots, Suture sponge 1 score |
| Cowan, 2021 | Randomized Controlled Trial | Surgeons, Fellows, residents | 34 | da Vinci SimNow | da Vinci Xi System dry lab | Analogous vesico-urethral anastomosis | Mixed level of experience and levels of training | Metrics automatically collected by the systems used | Total task time, Average ICA, Path length Path length, Camera distance, Out of view Unnecessary |
| Valdis, 2015 | Randomized Controlled Trial | Surgical trainees | 39 | dVSS | No additional Training | robotic ITA harvest and mitral annuloplasty task | Trainees with less than 10 hours of robotic surgical simulator experience | Videos recorded and assessed by 1 expert surgeon | Dissection scores, GEARS |
| Lerner, 2010 | Prospective Study | Medical students and interns | 22 | MdVT | Dry Lab | Peg board, patern cutting, letter board, string running, knot tying | No previous robotic experience | Automatically collected by VR and robot systems | Time to complete and accuracy |
| Hung, 2012 | Prospective randomized study | Surgical trainees | 24 | dVSS | No additional Training | pre and post training bowel resection, cystomy and repair, and partial | Robotic surgery trainees that completed less than 10 robotic cases | Performance recorded and reviewed by 3 expert surgeons | Time to complete, GOALS |
| Vargas, 2017 | Randomized Controlled Trial | Medical Students | 35 | dVSS | No additional Training | cystotomy closure on a live porcine model | Novice surgeons | Videos recorded and assessed by 4 expert surgeons | Time to complete, GEARS |
| Stegeman, 2013 | Cross-over Randomized Controlled Trial | Medical students, Residents, Fellows, Surgeons | 53 | FSRS | No additional Training | ball placement, suture pass, and fourth arm manipulation tasks | Limited robotic surgical experience | Videos recorded and assessed by 2 trained assessors | Time to complete, number of cemera and clutch movements, task errors, movement outside field of |
| Amirian, 2014 | Prospective randomized study | Medical Students | 26 | dVSS | Robotic Surgical System Dry lab | Penrose drain suture evaluation | Medical students | Procedure timed and penrose drain pulled on to determine quality of the suture | Time, accuracy, Gap |
| Cho, 2013 | Prospective Trial | Surgeons | 12 | mDVT | No additional Training | Automatic metrics recorded from simulator | Surgeons with similar laproscopic surgical experience | Performance metrics automatically recorded | Time, Number of errors, DV index |
| Bechtolsheim, 2023 | Prospective randomized study | Medical Students and Residents | 87 | dVSS | The DaVinci Xi surgical system Dry lab | Flap task, Precise Cut task, Dissection task, and Suture and Knot task | Medical students and First year residents without any formal robotic surgery experience | Peak force measured by device | Tissue Handeling, Time to Complete, Occurence of predefined errors |
| Chowriappa, 2014 | Randomized Controlled Trial | Residents and Fellows | 52 | HoST-AR | Watched 4 videos seeing surgery performed | Urethrovesical Anastomosis Performance on the daVinci Surgical System on an | Minimal or no robotic console experience of less than 25 hours | Performance video recorded and analyzed by experts | GEARS, NASA index |
| Satava, 2020 | Randomized Controlled Trial | Residents, fellows, attending surgeons | 99 | DVSS | dV-Trainer, no additional training, and dome | 5 tasks on an avian tissue model | Surgical robotics novices | videos recorded and reviewed by 2 blinded raters | Time to complete and numbers of errors |
| Whitehurst, 2015 | Prospective randomized study | Residents, fellows, attending surgeons | 20 | MdVT | da Vinci Xi System dry lab | cystotomy closure task in a swine model | Minimal exposure to Robotic surgical training | Video recorded and assessed by a group of blinded expert surgeons | GEARS, Time to complete, hand motion, |

**Table 2.** Overview of extended reality simulators available. NA = Not Available

| Name of simulator | Type | Type of Surgery | Methods | Haptics | Cost |
|---|---|---|---|---|---|
| da Vinci Skills Simulator [11] | VR/AR | General Robotic Surgery | Console with access to the physical surgical robot. Proficiency scores | No | $127,000 (console) $80,000 (w/o console) |
| dV-Trainer with Xperience Unit [12] | VR/AR | Cancerous Tumor Removal | Supervisor supervision w/o interruption, AR videos/interaction, Team training | Yes | $110,000 |
| Robotix Mentor [13] | VR/AR | Hemicolectomy, Prostatectomy | VR Procedures, Surgical console, Supervision console, access to physical surgical robot | No | $137,000 |
| Robotic Surgery Simulator [14] | VR/AR | Prostatectom, Hysterectomy | Video Modules, Force feedback, 3D viewer with monitor | Yes | $120,000 |
| ProMIS Simulator [15] | Hybrid | Laparoscopic Surgery | Laparoscopic Tools, Tactile Tasks, Objective proficiency scores. | Yes | $50,000 |
| Simsurgery Educational Platform [16], [17] | VR | Laparoscopic Surgery | General surgical robot simulator, Time-sensitive tasks | No | $62,000 |
| Robossis Surgical Simulator [18] | VR | Femur Fractures | Robossis System Femur Fracture Alignment | Yes | NA |
| Microsoft HoloLens Simulation [19] | MR | Open Liver Surgery, Laparoscopic Surgeries | Microsoft HoloLens 2, Unity AR Software, da Vinci Research Kit (DVRK) | No | $3,500 (HoloLens 2), $250 (DVRK), $180/Month |
| Simendo [20] | VR-based | Urology | Robotic urology, laparoscopic & robotic techniques | No | NA |
| Senhance Simulator [21] | VR-based | General Robotic Surgery | Senhance system procedures, haptic feedback and eye-tracking | Yes | NA |
| Versius Trainer [22] | VR-based | General Robotic Surgery | Designed for the Versus robotic system, general surgery and specialties | Yes | NA |
| FlexVR by Mimic Technologies [23] | VR-based | Flexible Robotic Surgery (i.e. NOTES) | Flexible robotic system simulation with procedural walkthroughs | No | $1,995 /Month |
| NeuroTouch [24] | VR-based | Robotic Neurosurgery | VR with tactile feedback, robotic brain surgery simulation | Yes | NA |

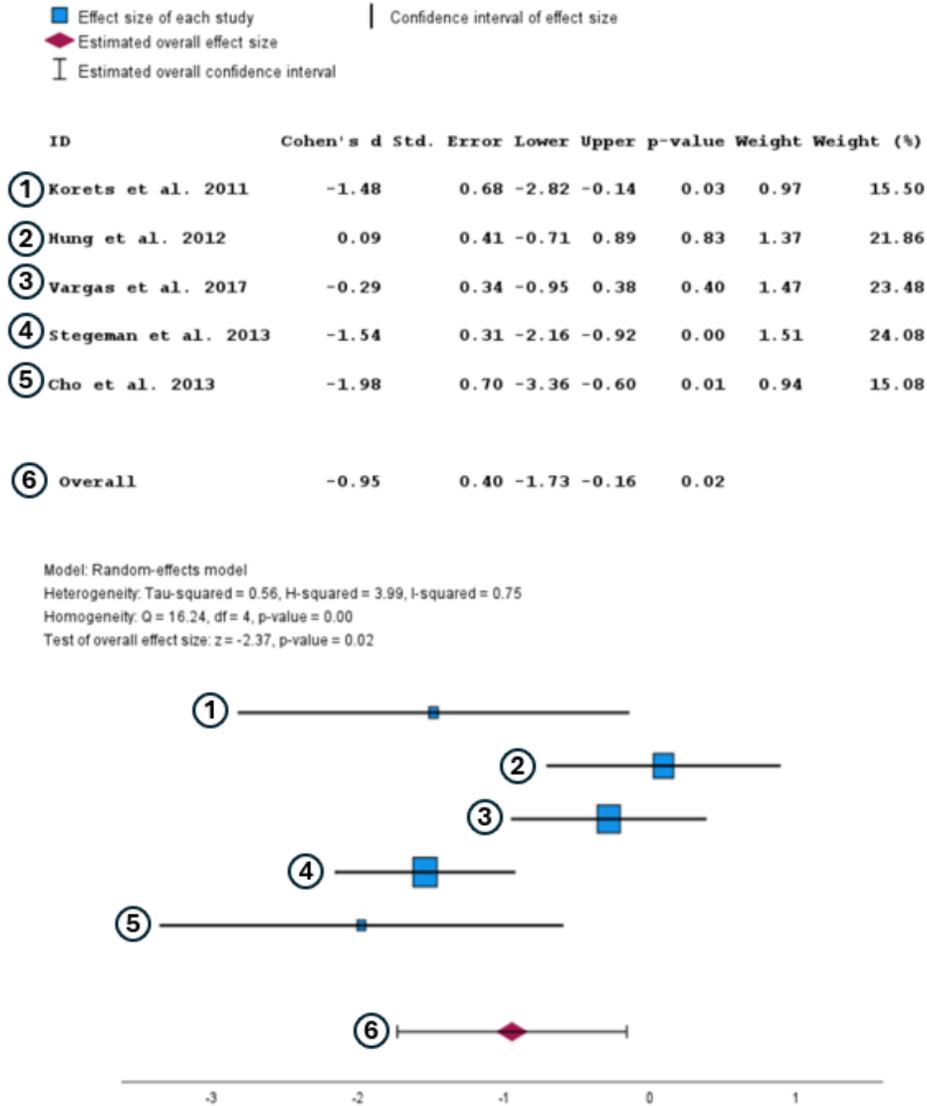

**Figure 2.** Forest plot comparing the effect size of VR training to no additional training on time to complete after completing a training regimen. [25], [26], [27], [28], [29]

  A random effects analysis of the GEARS scores showed that participants with extended reality training had higher GEARS scores than participants without any additional training. With the training having a large effect on the scores (Cohen's d=0.75, 95% CI [-0.48, 2.05]). However, this result is not statistically significant (P=0.23) (Figure 3).

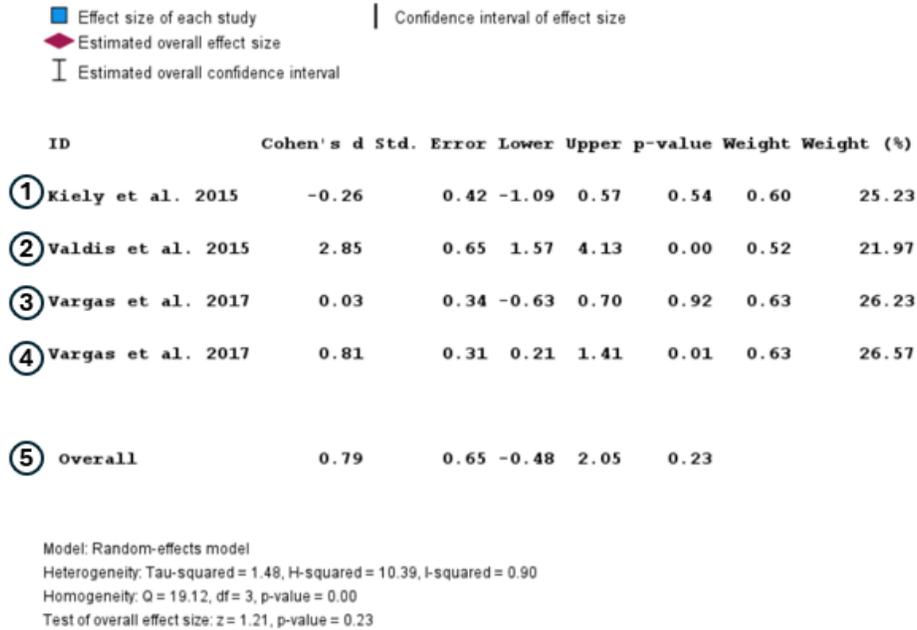

**Figure 3.** Forest plot comparing the effect size of VR training to no additional training on GEARS scores after completing a training regimen. [27], [30], [31]

*3.4 XR vs. Dry Lab*

To determine if using extended reality training leads to similar proficiencies in robotic surgeries as one of the most used training methods, the authors compared participants that were trained with extended reality simulators to participants that used inanimate, dry lab simulators.

A random effects analysis of time to complete showed a moderate, not statistically significant increase in time to complete for participants trained in virtual reality compared to participants trained in a dry lab (Cohen's d=0.65, 95% CI [-0.22,1.52], p=0.14) (Figure 4).

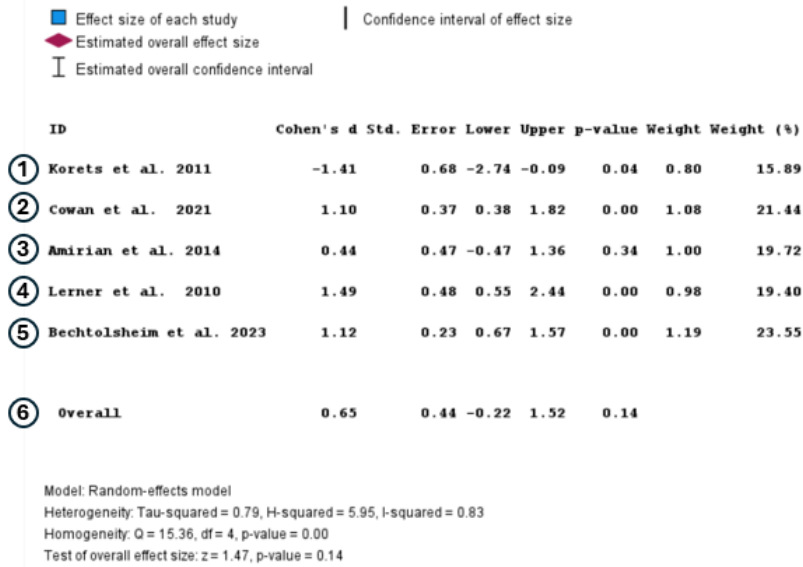

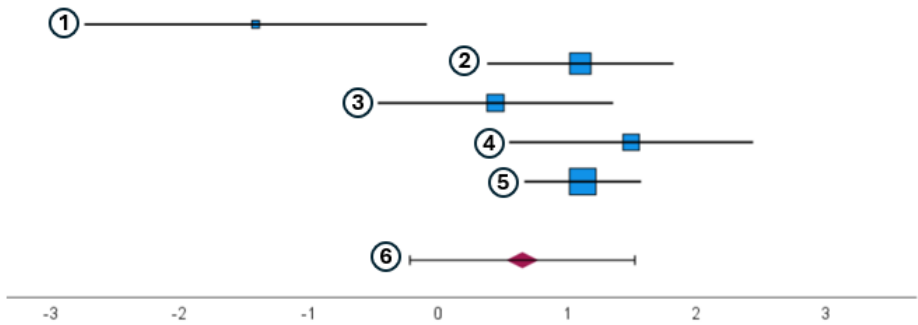

**Figure 4.** Forest plot comparing the effect size of VR training to dry lab training on time to complete after completing a training regimen. [25], [32], [33], [34], [35]

A random effects analysis of the GEARS scores showed that participants with Extended reality had lower GEARS scores than those that were trained in extended reality. (Cohen's d=-1.09, 95% CI [-3.17,0.98]). However, this result is not statistically significant (P=0.30) (Figure 5).

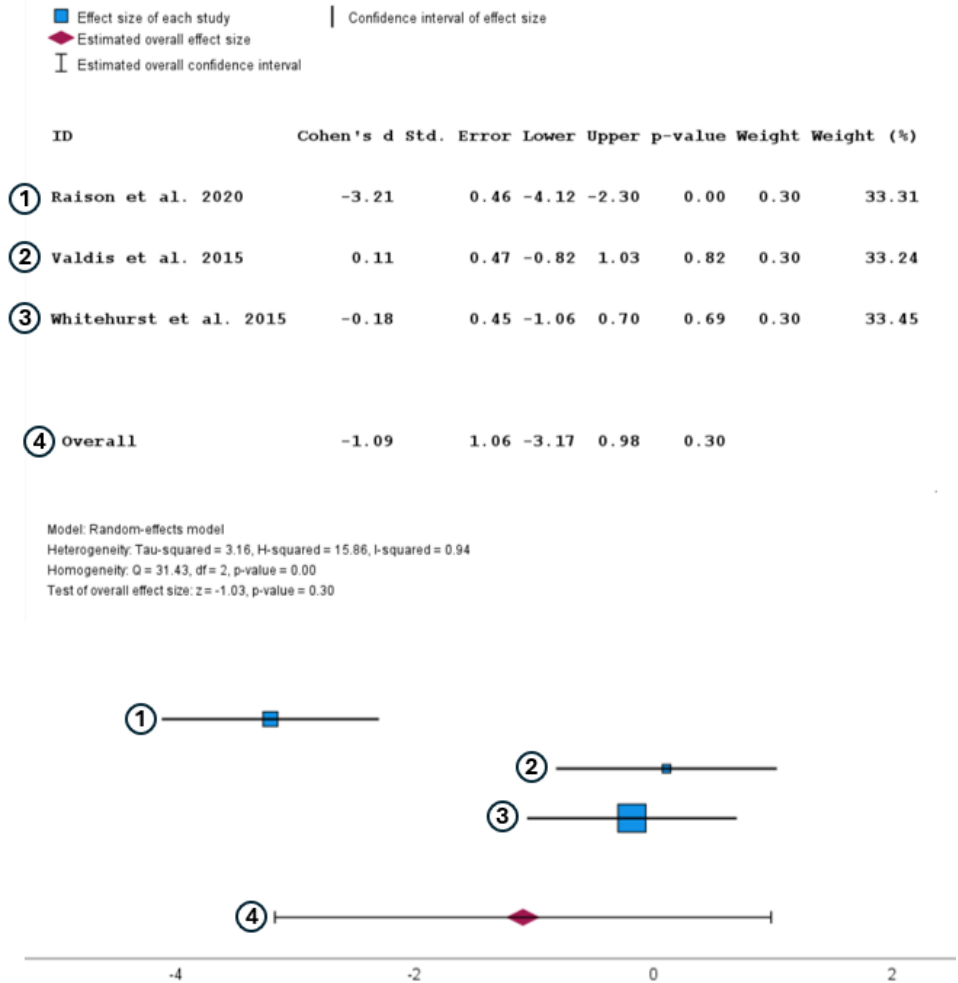

**Figure 5.** Forest plot comparing the effect size of VR training to dry lab training on GEARS Score after completing a training regimen. [31], [36], [37]

*3.5 GEARS score t-test*

  We performed an independent samples t-test to compare the mean GEARS score between VR and No Training groups (Table 3). Each group satisfied conditions for an independent samples t-test [11], as they were large in number (63 vs. 60 participants), worked in independent environments, and had highly homogenous standard deviations. The independent samples t-test yielded an effect size strongly in favor of the VR trained group, namely 0.964 (95% CI: 0.589 to 1.336). The threshold for a large effect size is 0.8 [13]. The mean GEARS score also crudely favored the VR trained group at 16.73, compared to no training at 13.91. The discrepancy is nearly three points. An independent samples t-test was also used to compare the mean GEARS score between VR and Dry Lab groups. Each group satisfied conditions for an independent samples t-test [11], as they were large in number (44 vs. 37 participants), worked in independent environments, and had highly homogenous standard deviations. The independent samples t-test yielded an inconclusive effect size of -0.093. This effect size rests in a 95% confidence interval of -0.530 to 0.345.

**Table 3.** Summary of independent sample t-tests

| Group | Mean±SD XR | Mean±SD Dry lab Training | Difference in Means | Cohen's d | P value |
|---|---|---|---|---|---|
| XR vs. No Training | 16.73±2.8, n=63 | 13.91±2.95, n=60 | 2.82±0.53 [95% CI 1.78, 3.86] | 0.964 [95% CI 0.589, 1.336] | <0.001 |
| XR vs. Dry lab | 13.43±6.7, n=44 | 14.08±7.33, n=37 | -0.65±1.56 | -0.093 [95% CI -0.530, 0.345] | 0.34 |

### 3.6 Risk of Bias and GRADE analysis

Each study was evaluated for an overall risk of bias using Cochrane's Risk of Bias handbook and tool by 2 authors independently [14]. A Robbins-I analysis was used due to the articles in this study including randomized controlled trails and prospective trials. 12 of the articles were found to have a low risk of bias while 3 of the articles were found to have a moderate risk of bias. Each domain of Robbins-I is represented in Figure 6 and Figure 7.

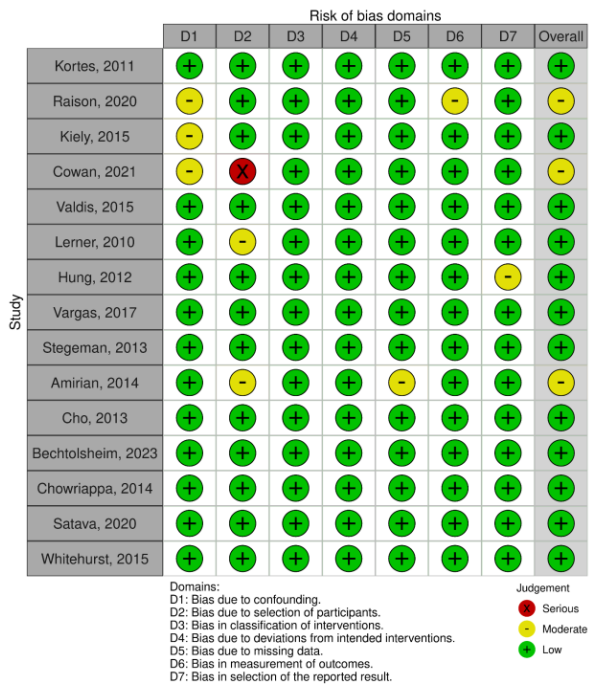

**Figure 6.** Traffic light plot portraying the Robbins-I assessment.

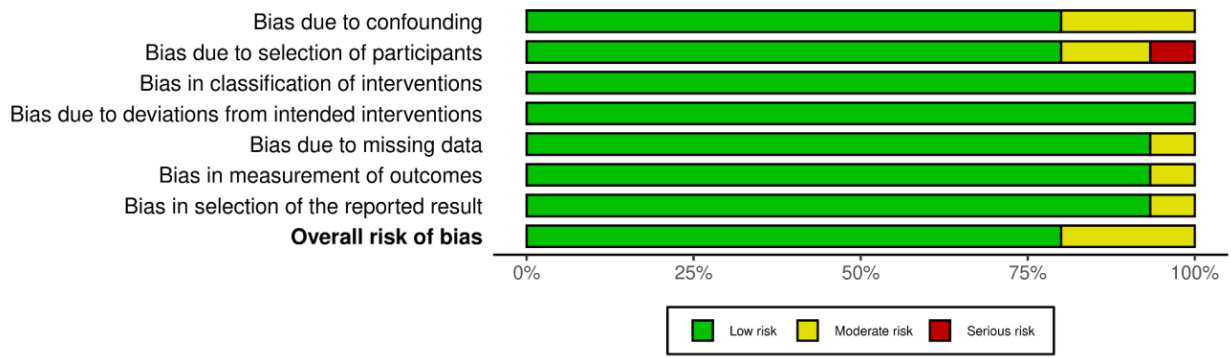

**Fig 7.** Robbin I summary plot

To determine the certainty of the evidence in this study, a GRADE analysis was performed by two independent authors. The analysis revealed the studies in this review are a moderate level of evidence (Table 4).

**Table 4**. Overall GRADE analysis of included articles

| Author | Year | Study Design | Intervention | Risk of Bias | Inconsistency | Indirectness | Imprecision | Grade |
|---|---|---|---|---|---|---|---|---|
| Korets et al [25] | 2011 | Randomized Controlled Trial | MdVT | Not Serious | Serious | Not serious | Serious | Moderate |
| Raison et al. [36] | 2020 | Prospective Study | dVSS | Serious | Serious | Not serious | Not Serious | Moderate |
| Kiely et al. [30] | 2015 | Randomized Controlled Trial | dVSS | Not Serious | Serious | Not serious | Not serious | Moderate |
| Cowan et al. [32] | 2021 | Randomized Controlled Trial | da Vinci SimNow | Serious | Serious | Not serious | Not Serious | Moderate |
| Valdis et al. [31] | 2015 | Randomized Controlled Trial | dVSS | Not Serious | Serious | Not serious | Not Serious | Moderate |
| Lerner et al. [34] | 2010 | Prospective Study | MdVT | Not Serious | Serious | Not serious | Not Serious | Moderate |
| Hung et al. [26] | 2012 | Prospective randomized study | dVSS | Not Serious | Serious | Not serious | Not Serious | Moderate |
| Vargas et al. [27] | 2017 | Randomized Controlled Trial | dVSS | Not Serious | Serious | Serious | Not Serious | Moderate |
| Stegeman et al. [28] | 2013 | Cross-over Randomized Controlled Trial | FSRS | Not Serious | Serious | Not serious | Not Serious | Moderate |
| Amirian et al. [33] | 2014 | Prospective randomized study | dVSS | Serious | Serious | Not serious | Not Serious | Moderate |
| Cho et al. [29] | 2013 | Prospective Trial | mDVT | Serious | Serious | Not serious | Serious | Moderate |
| Bechtolsheim et al. [35] | 2023 | Prospective randomized study | dVSS | Not Serious | Serious | Not serious | Not Serious | Moderate |
| Chowriappa et al. [8] | 2014 | Randomized Controlled Trial | HoST-AR | Not Serious | Serious | Not serious | Not Serious | Moderate |
| Satava et al. [38] | 2020 | Randomized Controlled Trial | DVSS | Not Serious | Serious | Not serious | Not Serious | Moderate |
| Whitehurst et al. [37] | 2015 | Prospective randomized study | MdVT | Not Serious | Serious | Not serious | Not Serious | Moderate |

## 4. Discussion
*4.1 Outcomes*

This systematic review and meta-analysis uses the most up to date randomized controlled and prospective trials to determine if extended reality simulators can be used to effectively train novice surgeons on how to use complex surgical robots. The summary of the meta-analyses completed in this study can be found in table 5. Participants who received XR training completed tasks in significantly less time than participants with no training. The effect size of -0.95 is considered large, per Cohen's d. The 95% confidence interval rested between -1.73 and -0.16, decidedly in favor of XR training. This means that surgeons trained with XR were able to complete tasks significantly quicker than those without training. The XR trained group also had a significantly higher average GEARS score (16.73) compared to the no training group (13.91). The margin of GEARS indicates better technical skills acquisition via XR training. The effect size of GEARS scores between XR and No Training groups was strongly in favor of XR training (0.964, 95% CI: 0.589 to 1.336), which is considered a large effect size. These results demonstrate that XR simulators can be used as an effective way to train surgeons compared to just having the surgeons become familiar with the robot before use.

**Table 5**. Summary of Meta-analysis results

| Comparison | Cohen's D | Confidence Intervals | P value |
|---|---|---|---|
| Extended reality vs. No training Time to complete | -0.95 | -1.73, -0.16 | 0.02 |
| Extended reality vs. No training GEARS Score | 0.75 | -0.48, 2.05 | 0.2 |
| Extended reality vs. Dry lab Time to complete | 0.65 | -0.22, 1.52 | 0.14 |
| Extended reality vs. Dry lab GEARS Score | -1.09 | -3.17, 0.98 | 0.30 |

This review also compared how effectively XR simulators trained novice surgeons compared to the more commonly used training method of inanimate models (dry labs). Participants who received XR training completed tasks slower than participants with dry lab training. This effect size is considered moderate with Cohen's d of 0.65. However, this result is not statistically significant with a p value of 0.14 and a 95% confidence interval that rests between -0.22 and 1.52, indicating that the results do not favor dry labs over ER. The GEARS scores of the ER group (13.43) and dry lab group (14.08) are also inconclusive with an effect size of -0.093, suggesting no significant difference in the quality of performance of participants in both groups. The wide confidence interval for the GEARS score effect size between XR and Dry Lab training (-0.530 to 0.345) indicates a range from strong favor for Dry Lab to moderate favor for XR training. These results indicate that XR training and dry lab training are comparable training methods and lead to similar results for surgeons.

We determined that XR training can be used as a model to train novice surgeons on surgical robots, due to it greatly improving the skills compared to participants with no training and having no statistically significant differences between XR and dry lab participants. The differences between the dry lab training and extended reality training could be due to many factors, including the dry lab participants having more experience using the surgical robot or the simulator not being realistic enough. However, with continued advances in virtual reality technology including better imaging and improved haptic feedback [39], the ability of surgical simulators to improve compared to dry lab simulators is massive. On top of the ability to continue to improve simulators, virtual reality offers some advantages over dry lab. With simulators, participants can retry a task as many times as they want, without needing to reset anything physically, as well as are able to have as many different situations as has been made by the simulator, while a physical model might be limited in the number of variations a participant can have. Finally, the cost of simulators is much less than using surgical robots to train surgeons. Not only are simulators less expensive than a fully trained surgical robot, but the cost of not being able to use the surgical robot for surgeries while training is occurring is also a loss. Due to this, dry labs are one of the most used training methods for novice surgeons right now. The development and improvement of surgical simulators should be continued to allow for better training in the future.

*4.2 Limitations*

The largest challenge with this review is the unstandardized way participants were assessed after being trained. Most studies used different organ systems, or models which lead to a wide variety of results in the time to complete, as each final task had a different standard of time to complete. The other risk of bias is the participant selection. Different studies used participants from a wide range of medical experience, leading to a major difference in the general skill level of the participants, even if they were all inexperienced with surgical robotics. Finally, while many studies used the same simulators, various training simulators with different durations of training were included in this review as well. This variation in training could lead to variation in different margins of time and different GEARS results.

## 5. Conclusion

RAS is particularly demanding to learn. However, through advances in virtual reality, augmented reality, and mixed reality, novice surgeons can practice these challenging surgeries without putting patients at risk. Many practice modalities have been created for this purpose, including animal models and inanimate models. However, extended reality simulators allow for an increase in variability in situation tasks and repeatability that these other models do not offer. This study is the first to compare surgical performance of different training modalities to extended reality post training and supports the use of implementing extended reality into the training of surgical residents.

## Acknowledgements

This research was supported by NIH Grant R01EB036365 and NSF Grants 2141099 and 2226489. Conflict of interest: Mohammad Abedin-Nasab is the founder at Robossis.